\documentclass[numberedappendix,apj,twocolumn]{emulateapj}

\usepackage[bookmarks,bookmarksnumbered,colorlinks=true, citecolor=blue, linkcolor=black,breaklinks]{hyperref}
\usepackage{epstopdf}

\newcommand{\like}{{\cal L}}

\newcommand{\fluxUnit}{erg~s$^{-1}$cm$^{-2}$}

\shorttitle{Spectroscopic Redshift for a Luminous $z=7.73$ Galaxy }
\shortauthors{Oesch et al.}

\submitted{Draft version \today}

\begin{document}

\title{A Spectroscopic Redshift Measurement for a Luminous Lyman Break Galaxy \\ at $z=7.730$ Using Keck/MOSFIRE}

\author{P. A. Oesch\altaffilmark{1,2},
P. G. van Dokkum\altaffilmark{2}, 
G. D. Illingworth\altaffilmark{3},
R. J. Bouwens\altaffilmark{4},
I. Momcheva\altaffilmark{2},
B. Holden\altaffilmark{3},  \\
G. W. Roberts-Borsani\altaffilmark{4,5},
R. Smit\altaffilmark{6},
M. Franx\altaffilmark{4}, 
I. Labb\'{e}\altaffilmark{4},
V. Gonz\'{a}lez\altaffilmark{7},
D. Magee\altaffilmark{3}
}

\altaffiltext{1}{Yale Center for Astronomy and Astrophysics, Physics Department, New Haven, CT 06520, USA; pascal.oesch@yale.edu}
\altaffiltext{2}{Department of Astronomy, Yale University, New Haven, CT 06520, USA}
\altaffiltext{3}{UCO/Lick Observatory, University of California, Santa Cruz, CA 95064, USA}
\altaffiltext{4}{Leiden Observatory, Leiden University, NL-2300 RA Leiden, The Netherlands}
\altaffiltext{5}{Department of Physics and Astronomy, University College London, Gower Street, London WC1E 6BT, UK}
\altaffiltext{6}{Centre for Extragalactic Astronomy, Department of Physics, Durham University, South Road, Durham DH1 3LE, UK}
\altaffiltext{7}{University of California, Riverside, 900 University Ave, Riverside, CA 92507, USA}

\begin{abstract}
We present a spectroscopic redshift measurement of a very bright Lyman break galaxy at $z=7.7302\pm0.0006$ using Keck/MOSFIRE. 
The source was pre-selected photometrically in the EGS field as a robust $z\sim8$ candidate with $H=25.0$ mag based on optical non-detections and a very red Spitzer/IRAC [3.6]$-$[4.5] broad-band color driven by high equivalent width [\ion{O}{3}]+H$\beta$ line emission. The Ly$\alpha$ line is reliably detected at $6.1\sigma$ and shows an asymmetric profile as expected for a galaxy embedded in a relatively neutral inter-galactic medium near the Planck peak of cosmic reionization. The line has a rest-frame equivalent width of $EW_0=21\pm4$~\AA\ and is extended with $V_\mathrm{FWHM}=360^{+90}_{-70}$ km~s$^{-1}$. The source is perhaps the brightest and most massive $z\sim8$ Lyman break galaxy in the full CANDELS and BoRG/HIPPIES surveys, having assembled already $10^{9.9\pm0.2}$ M$_\odot$ of stars at only 650 Myr after the Big Bang. The spectroscopic redshift measurement sets a new redshift record for galaxies. This enables reliable constraints on the stellar mass, star-formation rate, formation epoch, as well as combined [\ion{O}{3}]+H$\beta$ line equivalent widths. The redshift confirms that the IRAC [4.5] photometry is very likely dominated by line emission with $EW_0$([\ion{O}{3}]+H$\beta$)$= 720_{-150}^{+180}$ \AA. This detection thus adds to the evidence that extreme rest-frame optical emission lines are a ubiquitous feature of early galaxies promising very efficient spectroscopic follow-up in the future with infrared spectroscopy using JWST and, later, ELTs.
\end{abstract}


\vspace*{0.4truecm}

\section{Introduction}

The spectroscopic confirmation and characterization of galaxy candidates within the cosmic reionization epoch has been a major challenge for observational extragalactic astronomy for the last few years. Recently, large samples of several hundred galaxy candidates have been identified at $z\sim7-11$ thanks to the exceptional near-infrared sensitivity of the WFC3/IR camera onboard the Hubble Space Telescope \citep[$HST$; e.g.,][]{Bouwens11c,Bouwens14,Schenker13,McLure13,Oesch12a,Oesch14,Finkelstein14}. However, despite this unprecedented target sample, very little progress has been made in spectroscopically confirming galaxies in the cosmic reionization epoch. Currently, only a handful of normal galaxies have reliably measured redshifts at $z>7$ \citep[see, e.g.,][]{Vanzella11,Pentericci11,Ono12,Schenker12,Shibuya12,Finkelstein13}, with most spectroscopic surveys being unsuccessful or only resulting in uncertain candidate lines \citep[e.g.][]{Treu13,Jiang13a,Tilvi14,Caruana14,Faisst14,Vanzella14,Schenker14}.

Current studies at $z>6$ thus rely on photometric samples, with selection criteria and photometric redshifts that are somewhat uncertain due to the difficulty of establishing reliable priors of potential contaminant populations at lower redshift.
Spectroscopic follow-up is therefore particularly important for the very rare galaxies at the bright end of the UV luminosity and mass functions where any contamination has a very large impact.

The low success rate of spectroscopic follow-up surveys is likely caused by a decreased fraction of galaxies showing strong Ly$\alpha$ emission due to an increased neutral fraction in the inter-galactic medium (IGM) at $z>6$ \citep[][]{Stark10,Stark11,Treu13,Schenker12,Schenker14,Pentericci14}. While Ly$\alpha$ is the primary spectral feature for spectroscopic confirmation of high-redshift candidates, new surveys targeting the Lyman continuum break are underway using the WFC3/IR grism on the $HST$, or alternatively, weak UV lines may also be detectable from the ground \citep[][]{Stark14b}. 

Two recent successful Ly$\alpha$ detections of Lyman break selected galaxies at $z>7.2$ were published in \citet[][$z=7.213$]{Ono12} and \citet[][$z=7.508$]{Finkelstein13}. Both these sources are relatively bright with $H=25.2$, and $25.6$ mag and both show a significant flux excess in their IRAC photometry, which is consistent with extremely strong [\ion{O}{3}] line emission at $4.1-4.3~\mu$m. Such strong lines are characteristic of early star-forming galaxies, as evidenced by a clear increase in broad-band flux excess with redshift \citep[e.g.,][]{Schaerer09,Labbe13,Stark13,Gonzalez12b,Smit13}. Such excesses can be used to select relatively clean samples of star-forming galaxies at $z\sim6.6-6.9$ as well as $z\sim8$ \citep[e.g.][Roberts-Borsani et al., 2015, in prep.]{Smit14}.

In this $Letter$ we present a successful spectroscopic redshift measurement at $z\sim8$ using Keck/MOSFIRE of one of the brightest Lyman Break galaxies (LBGs) at that epoch. This galaxy was pre-selected as a high-priority target because of a very red [3.6]-[4.5] color, likely caused by strong [\ion{O}{3}] emission. 
Our target selection is summarized in Section \ref{sec:targets}, while Section \ref{sec:obs} outlines the spectroscopic observations, and our results are presented in Section \ref{sec:results}.  

Throughout this paper, we adopt $\Omega_M=0.3, \Omega_\Lambda=0.7, H_0=70$ kms$^{-1}$Mpc$^{-1}$, i.e. $h=0.7$, consistent with the measurements from Planck \citep{Planck2015}. Magnitudes are given in the AB system. % \citep{Oke83}. 

\begin{figure}[tbp]
	\centering	
	\includegraphics[width=\linewidth]{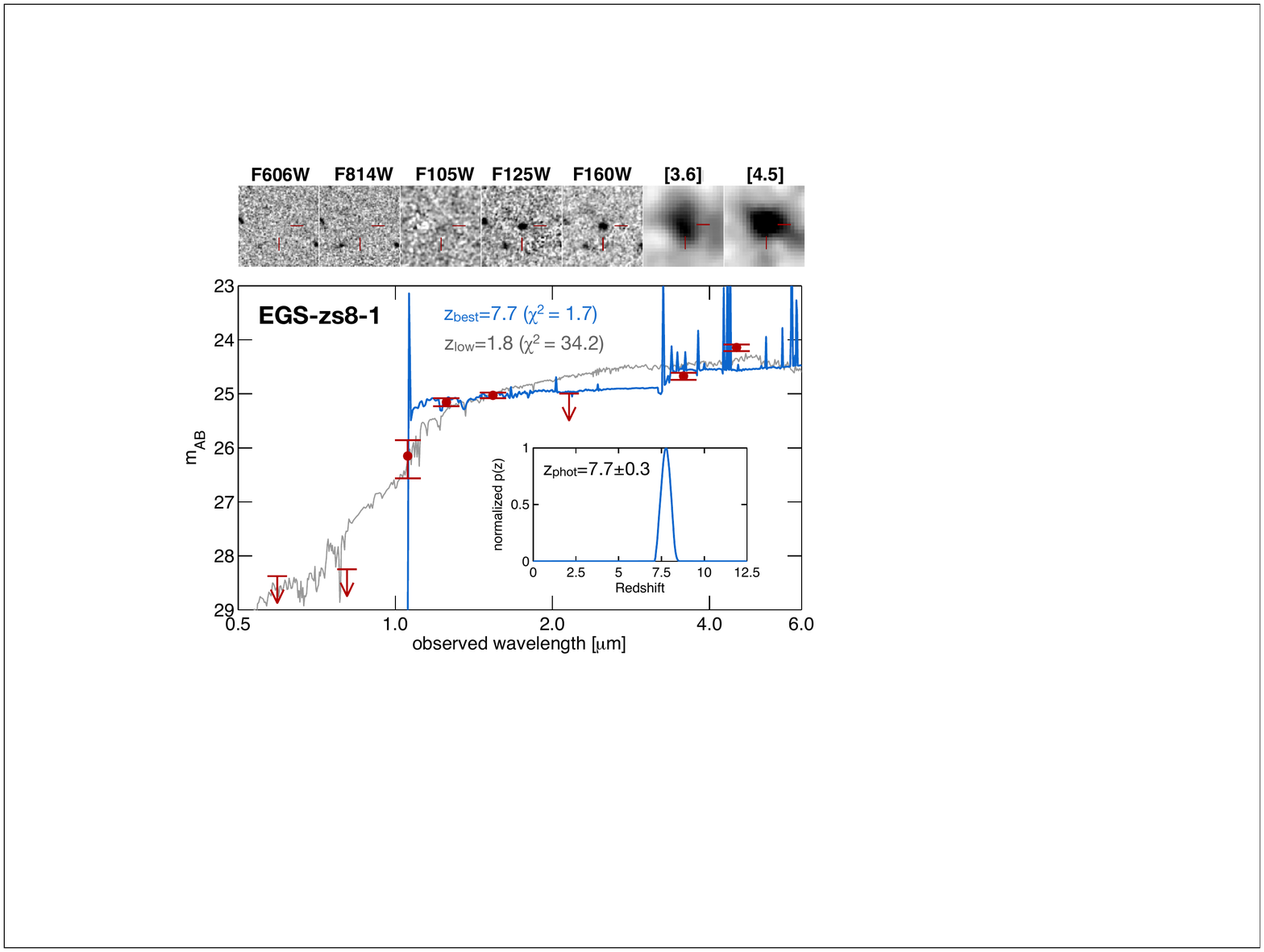}
  \caption{\textit{Top -- } Images showing a 5\arcsec$\times$5\arcsec\ region around our primary target galaxy EGS-zs8-1 in the HST and Spitzer/IRAC filters. These are, from left to right, $V_{606}$, $I_{814}$, $Y_{105}$, $J_{125}$, $H_{160}$, and 3.6 $\mu$m and 4.5 $\mu$m. 
  \textit{ Bottom --} The spectral energy distribution of EGS-zs8-1 based on fits to the HST+Spitzer+K-band photometry. Downward pointing vectors represent 2 $\sigma$ upper limits in non-detection bands. A significant flux excess in the IRAC 4.5 $\mu$m band is evident. Together with the strong spectral break, this constrains the photometric redshift to $z_{phot} = 7.7\pm0.3$, in excellent agreement with the spectroscopic measurement as shown later. The best fit low-redshift solution at $z\sim1.8$ is shown as a gray line for completeness. However, this SED has a likelihood of $<10^{-7}$ and is ruled out by the photometry. 
  }
	\label{fig:stamps}
\end{figure}

\section{Target Selection}
\label{sec:targets}

We briefly summarize our selection of a robust  $z\sim8$ LBG sample over the CANDELS fields using extreme IRAC photometry. For more details see Roberts-Borsani et al. (2015, in prep.). 

The selection builds on \citet{Smit14}, who identify a sample of $z\sim6.8$ galaxies based on strong [\ion{O}{3}]$\lambda\lambda4959,5007$ plus H$\beta$ emission lines resulting in very blue [3.6]$-$[4.5] IRAC colors. As these lines shift into the IRAC 4.5 $\mu$m band, galaxies at $z\sim7$ to $z\sim9$ exhibit red [3.6]$-$[4.5] IRAC colors \citep[see also][]{Stark13,Labbe13,Bowler14}. 

We exploit the availability of deep Spitzer/IRAC photometry over the HST CANDELS-Wide fields to systematically search for bright galaxies with IRAC colors of [3.6]$-$[4.5]$>0.5$ mag in addition to a Ly$\alpha$ break (i.e., a non-detection at $<1~\mu$m), characteristic for $z>7$ galaxies. This resulted in two candidates with $H<25.1$ mag in the EGS field (see Roberts-Borsani et al., in prep.). Fortuitously, these two sources are $<6\arcmin$ from each other and can be targeted in a single MOSFIRE mask. 

Stamps and SED fits for one of these sources (EGS-zs8-1) are shown in Figure \ref{fig:stamps}. 
The F606W, F814W, F125W and F160W images come from the CANDELS survey \citep[][]{Grogin11}, while the IRAC images are from the SEDS survey \citep{Ashby13}. Also shown are WFC3/IR F105W observations that are fortuitously available over this source as a result of a separate follow-up program (GO:13792, PI: Bouwens). Despite its modest depth, this $Y_{105}$ image still provides a highly improved photometric redshift measurement by constraining the spectral break at $1~\mu$m. 

As seen from the figure, the source EGS-zs8-1 is only detected at $>1~\mu$m in the WFC3/IR imaging as well as in both IRAC 3.6 and 4.5 $\mu$m bands. The [3.6]$-$[4.5] color of this source is measured to be 0.53$\pm$0.09, i.e., at the edge of our IRAC color selection window ([3.6]$-$[4.5]$>0.5$).

\section{Observations}
\label{sec:obs}

\subsection{MOSFIRE Spectroscopy}

We use the Multi-Object Spectrometer for Infra-Red Exploration \citep[MOSFIRE;][]{McLean12} on the Keck 1 telescope for Y-band spectroscopy of our primary $z\sim8$ targets in the search for their Ly$\alpha$ emission lines. MOSFIRE offers efficient multiplex observations over a field of view of $\sim6\arcmin\times3\arcmin$ at a spectral resolution of $R\sim3000$ ($R=3500$ and $R=2850$ for a 0\farcs7 or 0\farcs9 slit, respectively).

Data over the EGS field were taken during three nights, 2014 April 18, 23, and  25. 
While the first night was essentially lost due to bad seeing and clouds, the remaining nights had better conditions with a median seeing of $1\arcsec$ and only few cirrus clouds during the last night.
A dome shutter break problem also led to some vignetting during the last 30 minutes of the April 25 night before we stopped observations early. 
In total, we obtained 2.0 hours of good quality Y-band spectroscopy during April 23 (Night 1), and 2.0 hours on April 25 (Night 2). 

Data were taken with 180 s exposures and AB dither offsets along the slit with $\pm1\arcsec$ and $\pm1.2\arcsec$, respectively. In night 2, we also increased the slit width from 0\farcs7 (as used in Night 1) to 0\farcs9 in anticipation of the slightly worse seeing forecast. 
During these nights we observed two masks with a total of eight $z\sim7-8$ candidate galaxies, in addition to lower redshift fillers.

\subsection{Data Reduction}

The data were reduced using a modified version of the public MOSFIRE reduction code DRP\footnote{https://code.google.com/p/mosfire/}. This pipeline produces 2D sky-subtracted, rectified, and wavelength-calibrated data for each slitlet with a spatial resolution of 0\farcs1799 per pixel and a dispersion of 1.086 \AA\ per pixel. 

Each of our mask contains one slitlet placed on a star for monitoring the sky transparency and seeing conditions of each exposure. We use this star to track the mask drift across the detector \citep[see, e.g.,][]{Kriek14}, which we find to be $\pm1.5$ pixels ($\pm$0\farcs27) and $\pm1$ pixel ($\pm$0\farcs18) during night 1 and 2, respectively. We separately reduce different batches of the data (of 30-45 min duration) to limit any S/N reduction caused by this drift, before shifting and stacking the data.

\begin{figure}[tbp]
	\centering	
	\includegraphics[width=\linewidth]{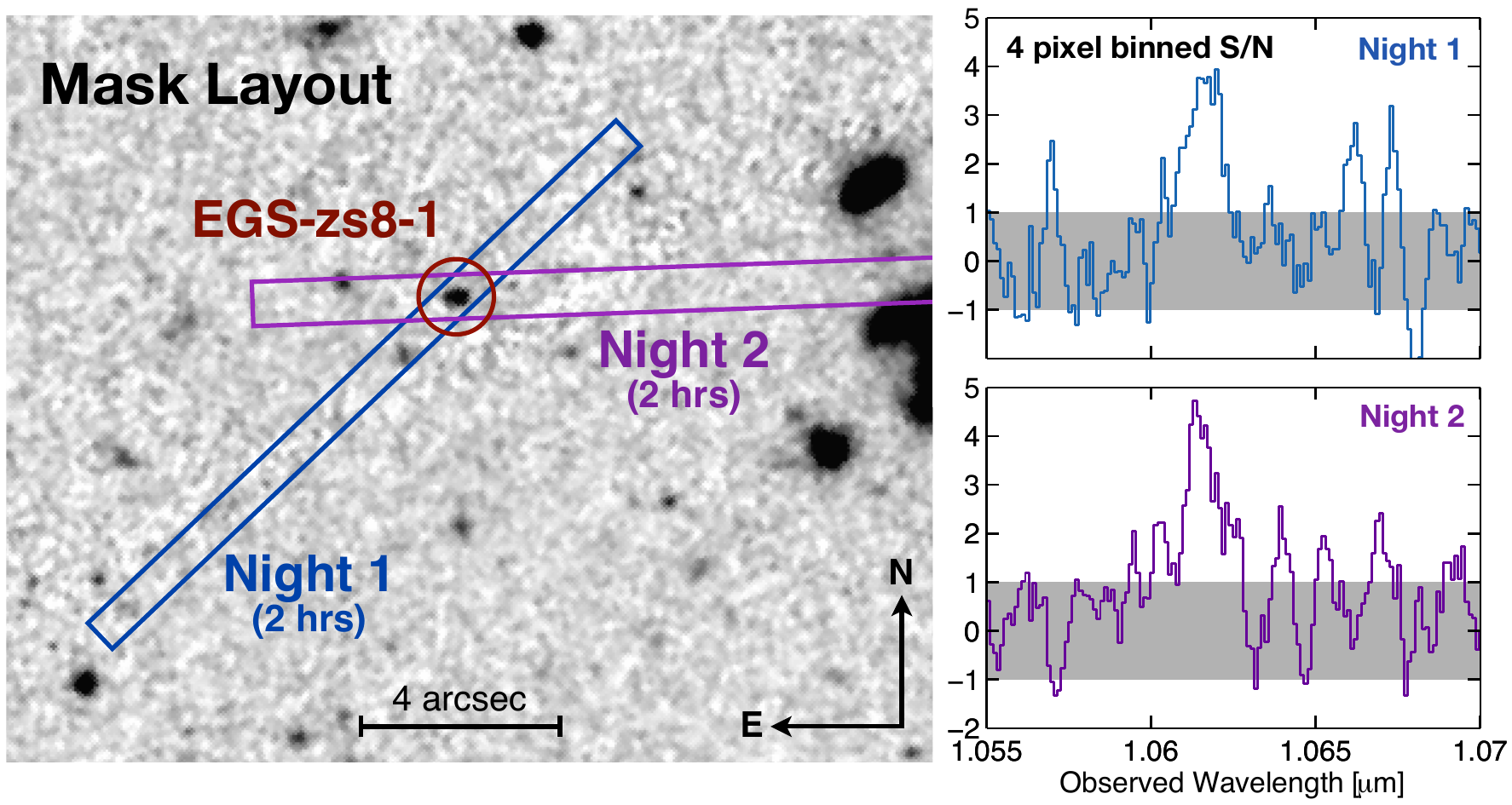}
  \caption{\textit{Left -- } Mask layout of the two nights of MOSFIRE Y-band observations of our primary target. These two nights provide two completely independent measurements of this galaxy at two different orientations as well as two different positions along different slitlets. This also allows us to exclude the possibility of contamination in the final stacked spectrum from the two faint neighboring galaxies present within 2\arcsec\ of the primary galaxy along the slits. 
  \textit{ Right -- } The signal-to-noise ratio around the detected emission line in the two independent 1D spectra of the two nights, averaged over a 4 pixel width ($\sim4$ \AA). A line is clearly detected at $>4\sigma$ independently in both 2 hr spectra from each night. We also checked the unrectified frames to ensure that the positive flux in the spectrum indeed originated from the expected position of the galaxy along the spectrum.
  }
	\label{fig:maskLayout}
\end{figure}

The masks for the two nights have different orientations (Fig \ref{fig:maskLayout}). The two independent data sets of the primary target thus add to the robustness of any detection. After creating the 2D spectra for the different masks, we applied the appropriate relative shift of the two 2D frames before stacking the observations of the two nights to our final 2D spectrum. 

Similarly, 1D spectra were extracted separately for each mask using an optimal extraction based on a profile determined by the slit star. The extracted 1D spectra were corrected for Galactic extinction and for telluric absorption using nearby A0 stars observed in the same night at similar airmass. The uncertainty in our optimally extracted 1D spectra was determined empirically from empty rows in the full, rectified 2D spectra of the mask. 

The absolute flux calibration was obtained from the slit stars by comparison of the spectra with the 3D-HST photometric catalogs \citep[][]{Skelton14}. An additional small correction was applied to account for the extension of individual sources in the slit mask by integrating the seeing-matched HST images over the slit and comparing with the slit loss of stellar sources.

\begin{figure}[tbp]
	\centering	
	\includegraphics[width=\linewidth]{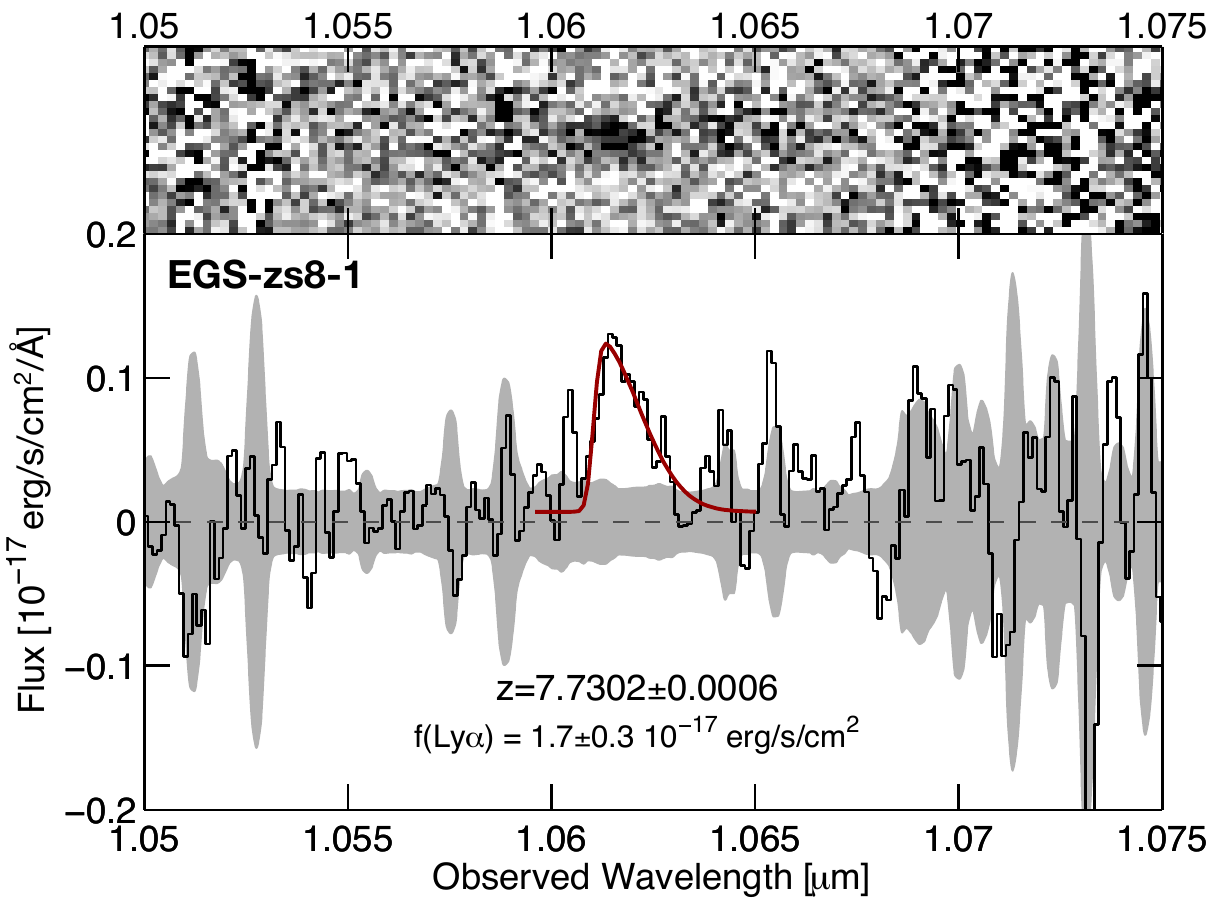}
  \caption{MOSFIRE spectra of EGS-zs8-1. The full 2D spectrum after 2-by-2 binning is shown in the top panel, while the optimally extracted 1D spectrum is shown on the bottom. The 1D spectrum was smoothed by a 3 pixel ($\sim3$ \AA) moving average filter for clarity. The gray shaded area represents the 1$\sigma$ flux uncertainty, while the dark red line shows the best-fit model. The line is quite extended in the wavelength direction and shows clear asymmetry with the expected shape typical for high-redshift Ly$\alpha$ lines. The spectroscopic redshift measurement is $z_\mathrm{spec} = 7.7302\pm0.0006$ in excellent agreement with the previously determined photometric redshift. Other line characteristics are summarized in Table \ref{tab:specsummary}. 
  }
	\label{fig:spectrum}
\end{figure}

\begin{deluxetable}{ll}
\tablecaption{Measurements of Galaxy EGS-zs8-1}
\tablecolumns{2}
\tablewidth{0.8\linewidth}

\startdata
\cutinhead{Target }\\[-3pt]
R.A. (J2000)  	&  14:20:34.89 \\
Dec (J2000) 		&  53:00:15.4\\
$H_{160}$    	&  25.03$\pm$0.05 \\
$M_{UV}$			&  $-22.06\pm0.05$ \\

\cutinhead{Emission Line}\\[-3pt]
$z_\mathrm{spec}$ 	& 7.7302$\pm$0.0006 \\
f(Ly$\alpha$) 		&  1.7$\pm$0.3$\times10^{-17}$ \fluxUnit \\
{L(Ly$\alpha$)} 		& 1.2$\pm$0.2$\times10^{43}$ erg~s$^{-1}$ \\
{EW$_0$(Ly$\alpha$)\tablenotemark{a}} & 21$\pm$4 \AA \\
$S_w$ 		&   15$\pm$6 \AA \\
{FWHM}\tablenotemark{b} 		&   13$\pm$3 \AA \\
$V_\mathrm{FWHM}$\tablenotemark{b}  &  360$_{-70}^{+90}$ km~s$^{-1}$ \\

\cutinhead{Physical Parameters\tablenotemark{c}}\\[-3pt]
$\log M_{gal}/M_\odot$  &  $9.9\pm$0.2 \\
$\log \mathrm{age}/$yr  & 8.0$\pm$0.5  \\
$\log \mathrm{SFR}/(M_\odot \mathrm{yr}^{-1})$ & $1.9\pm0.2$ \\
log SSFR			&  $-8.0\pm0.4$ \\
A$_\mathrm{UV}$  & 1.6 mag \\
UV slope $\beta$  & $-1.7\pm$0.1 \\[-5pt]

\enddata

\tablenotetext{a}{Not corrected for IGM absorption.}
\tablenotetext{b}{Derived from truncated Gaussian fit, corrected for instrumental broadening, but not for IGM absorption.}
\tablenotetext{c}{Based on SED fits \citep[see Sect 5;][]{Oesch14}. }
\label{tab:specsummary}
\end{deluxetable}

\section{Results}
\label{sec:results}

Out of the eight $z\sim7-8$ galaxy candidates, we detected a significant emission line (at $>5\sigma$) for only one source (EGS-zs8-1). This line is discussed in detail below.

\subsection{A Ly$\alpha$ Emission Line at $z=7.730$}

The spectra of our target source EGS-zs8-1 (see Table \ref{tab:specsummary} for summary of properties) revealed a significant emission line at the expected slit position in both masks independently (right panels Fig \ref{fig:maskLayout}). The full 4 hr stacked 2D and 1D spectra are shown in Figure \ref{fig:spectrum}, showing a line with a clear asymmetric profile, as expected for a Ly$\alpha$ line at high redshift ($z\gtrsim3$). Furthermore, it lies at the expected wavelength based on our photometric redshift estimate $z_\mathrm{phot} = 7.7\pm0.3$. We therefore interpret this line as Ly$\alpha$ (other possibilities are discussed in section \ref{sec:caveats}).

We fit the line using a Markov Chain Monte Carlo (MCMC) approach based on the \texttt{emcee} python library \citep[][]{emcee13}. Our model is based on a truncated Gaussian profile to account for the IGM absorption and includes the appropriate instrumental resolution. The model also includes the uncertainty on the background continuum level. The MCMC output provides full posterior PDFs and uncertainties for the redshift, line flux, significance, and line width. 

The line corresponds to a redshift of $z_{\mathrm{Ly}\alpha} = 7.7302\pm0.0006$, with a total luminosity of $L_{\mathrm{Ly}\alpha} =1.2\pm0.2\times10^{43}$ erg~s$^{-1}$, and a total detection significance of $6.1\sigma$. This is somewhat lower, but consistent with a simple estimate of 7.2$\sigma$ detection significance from integrating the 1D extracted pixel flux over the full extent of the line (i.e., not accounting for background continuum offsets).

Note that the redshift of the line is determined from our model of a truncated Gaussian profile and is thus corrected for instrumental resolution and the asymmetry arising from the IGM absorption. The peak of the observed line ($\lambda = 10616$ \AA) thus lies $\sim2.5$ \AA\ to the red of the actual determined redshift.

Given the brightness of the target galaxy, the detected line corresponds to a rest-frame equivalent width EW$_0=21\pm4$ \AA.  This is lower than the Ly$\alpha$ emitter criterion EW$_0>25$ \AA\ set in recent analyses that use the Ly$\alpha$ fraction among LBGs to constrain the reionization process \citep[e.g.][]{Stark11, Treu13}.

\subsection{Line Properties}

Different quantities of the detected line are tabulated in Table \ref{tab:specsummary}. In particular, we compute the weighted  skewness parameter, $S_w$ \citep{Kashikawa06} finding $S_w=15\pm6$ \AA. This puts the line above the 3 \AA\ limit found for emission lines at lower redshift (see also section \ref{sec:caveats}). 

The full-width-at-half-maximum (FWHM) of the line is quite broad with $\mathrm{FWHM}=13\pm3$ \AA, corresponding to a velocity width of $V_\mathrm{FWHM}=360_{-70}^{+90}$ km~s$^{-1}$. Our galaxy thus lies at the high end of the observed line width distribution for $z\sim5.7-6.6$ Ly$\alpha$ emitters \citep[e.g.][]{Ouchi10}, but is consistent with previous $z>7$ Ly$\alpha$ lines \citep[][]{Ono12}.

\subsection{Caveats}
\label{sec:caveats}

While the identification of the detected asymmetric emission line as Ly$\alpha$ is in excellent agreement with the expectation from the photometric redshift, we can not rule out other potential identifications. As pointed out in the previous section, \citet{Kashikawa06} find that weighted asymmetries $S_w>3$ \AA\ are not seen in lower redshift lines, but almost exclusively in Ly$\alpha$ of high-redshift galaxies. However, at the resolution of our spectra, the observed asymmetry is also consistent with an [\ion{O}{2}] line doublet in a high electron density environment, i.e., with a ratio of [\ion{O}{2}]$\lambda 3726$ / [\ion{O}{2}]$\lambda 3729 > 2$, and with a velocity dispersion of $\sigma_v\gtrsim100$ km~s$^{-1}$. 

If the observed line is an [\ion{O}{2}]$\lambda\lambda3726,3729$ doublet, the redshift of this galaxy would be $z_\mathrm{OII} = 1.85$. This is very close to the best low redshift SED fit shown in Figure \ref{fig:stamps}. However, that SED requires a strong spectral break caused by an old stellar population, for which no emission line would be expected. Additionally, the low-redshift solution can not explain the extremely red IRAC color (used to select this galaxy), and predicts significant detections in the ACS/F814W band, as well as in the ground-based WIRDS $K$-band image \citep[][]{Bielby12}. No such detections are present, however, resulting in a likelihood for such an SED of $\like<10^{-7}$ (see also Fig \ref{fig:stamps}). 
Thus all the evidence points to this line being Ly$\alpha$ at $z=7.73$.

\begin{figure}[tbp]
	\centering	
	\includegraphics[width=0.95\linewidth]{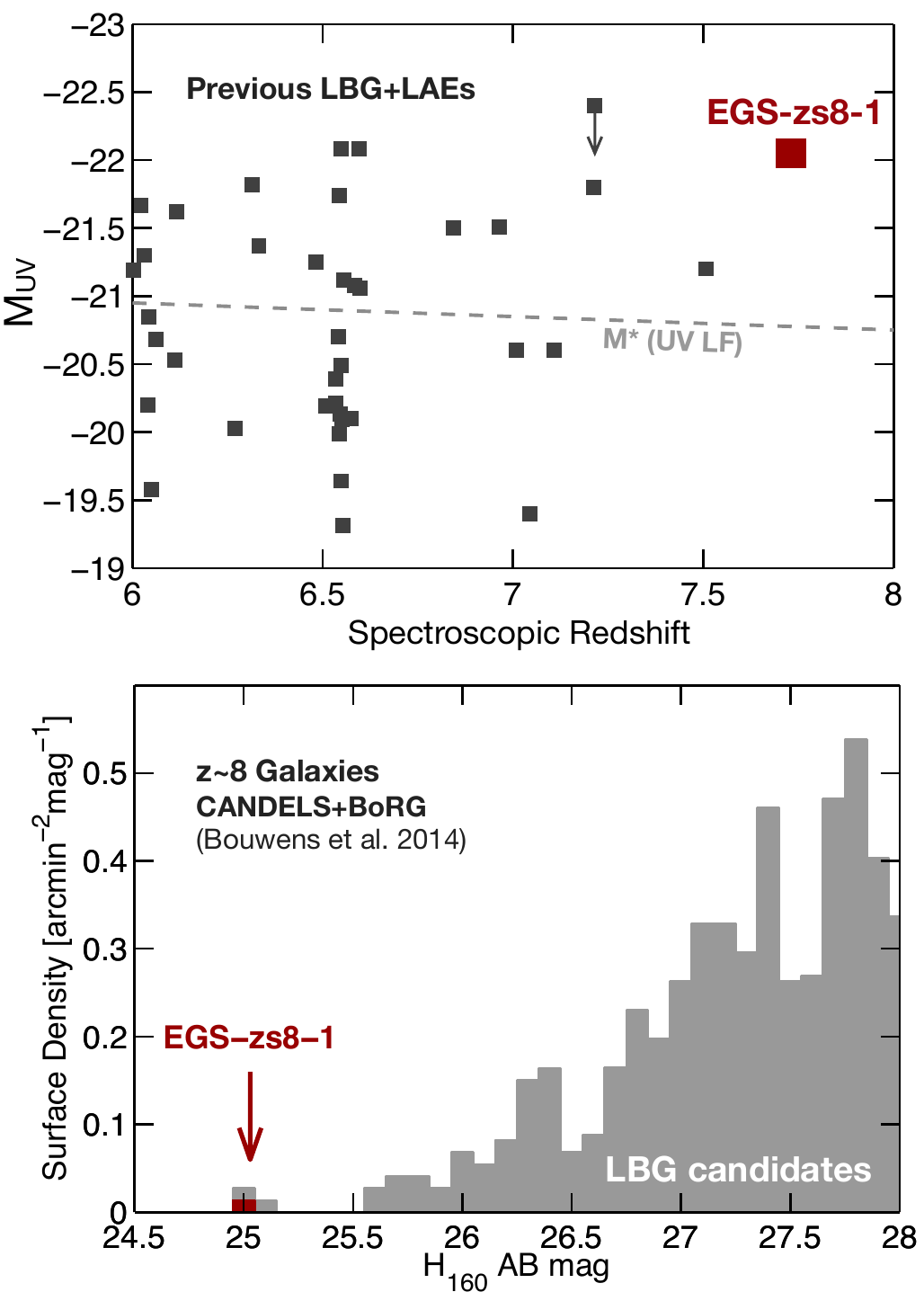}
  \caption{\textit{Top -- } UV absolute magnitudes of spectroscopically confirmed Lyman break galaxies and Ly$\alpha$ emitters in the cosmic reionization epoch, at $z>6$. Our target, EGS-zs8-1 (red square), represents the highest-redshift source and is the brightest galaxies currently confirmed. For reference, the gray dashed line shows the evolution of the characteristic magnitude $M_*$ of the UV LF \citep{Bouwens14}.
  The galaxies shown as black squares are assembled from a compilation from \citet{Jiang13b,Finkelstein13,Shibuya12,Ono12}, and \citet{Vanzella11}. 
  \textit{ Bottom -- } Surface density of the full sample of $z\sim8$ galaxies in the combined CANDELS and BoRG/HIPPIES fields \citep[][gray histogram]{Bouwens14}. EGS-zs8-1 is the brightest and also one of the most massive sources at these redshifts. Note that all three $z\sim8$ candidates with $H\sim25.0$ mag ($\sim0.5$ mag brighter than the rest) are identified in the CANDELS/WIDE survey area where ancillary $Y_{105}$ imaging is generally not available. Our spectroscopic confirmation is thus especially valuable.
  }
	\label{fig:fig4}
\end{figure}

%\newpage

\section{Discussion}

In this $Letter$ we used Keck/MOSFIRE to spectroscopically confirm the redshift of one of the brightest $z\sim8$ galaxies 
identified by \citet{Bouwens14} over the five CANDELS fields. Interestingly, this source is $\sim0.5$ mag brighter than any source identified in the wide-area BoRG and HIPPIES surveys \citep[e.g.][]{Trenti11,Yan11,Bradley12,Schmidt14a}.

As shown in Fig. \ref{fig:fig4}, with $z_\mathrm{spec} = 7.730$ and an absolute magnitude $M_\mathrm{UV} = -22.06\pm0.05$ the source EGS-zs8-1 is currently the most distant and brightest spectroscopically confirmed galaxy \citep[apart from a gamma ray burst at $z=8.2$;][]{Tanvir09,Salvaterra09}.
EGS-zs8-1 also populates the brightest bin of the recent \citet{Bouwens14} $z\sim8$ UV luminosity function (LF), which makes it an unusually rare
object. A spectroscopic confirmation of its high redshift is thus particularly valuable for proving the
existence of bright $H = 25.0$ mag galaxies at $z\sim8$ and for validating the bright end LF constraints.

An SED fit at the spectroscopic redshift of the source reveals a relatively high stellar mass $\log M/M_\odot = 9.9\pm0.2$, a star-formation rate of $\log \mathrm{SFR}/(M_\odot \mathrm{yr}^{-1}) = 1.9\pm0.2$, and a relatively young, but not extreme age of $\log \mathrm{age}/\mathrm{yr} = 8.0\pm0.5$ based on an apparent Balmer break between the WFC3/IR and the Spitzer photometry (see also Table \ref{tab:specsummary}). The corresponding formation redshift of this galaxy thus lies at $z_f=8.8$. For details on our SED fitting see, e.g., \citet{Oesch14}.

Interestingly, the source has a UV continuum slope of $\beta=-1.7\pm$0.1 (measured from the SED fit) and is consistent with considerable dust extinction, E(B$-$V)$=0.15$ mag. The detection of a significant Ly$\alpha$ emission line is not inconsistent, however, given
the complexities of line formation in such young galaxies. 
%This also highlights the need for high-$z$ LBG selections to ensure the inclusion of such relatively red sources 

These observations also allow us to reliably constrain the equivalent widths of the [\ion{O}{3}]+H$\beta$ emission lines in this galaxy based on its IRAC colors. At the spectroscopic redshift of the source, these lines are shifted in the 4.5 $\mu$m channel resulting in a color of [3.6]$-$[4.5]$=0.53\pm0.09$ mag. This is consistent with a combined rest-frame equivalent width of EW$_0$([\ion{O}{3}]+H$\beta$)$= 720_{-150}^{+180}$ \AA. 

Such a high equivalent width is in some tension with the inferred stellar population age of $\sim100$ Myr. However, it is completely consistent with the average EW$_0$([\ion{O}{3}]+H$\beta$) found for $z\sim7-8$ galaxies in previous work \citep[e.g.][]{Labbe13,Smit13,Smit14}, where such strong lines were found to be ubiquitous \citep[see also][]{Laporte14}.

The fact that these strong lines are seen in a significant fraction of the $z\sim7-8$ galaxies is at odds with the interpretation of extremely young galaxy ages of $<10$ Myr \citep[e.g.][]{Finkelstein13}. Nevertheless, very stochastic star-formation may explain some of the line strength. Instead, it is possible that an evolution in the ionization properties of early galaxy populations may be causing stronger emission lines with more extreme line ratios [\ion{O}{3}]/H$\beta$ as is observed at $z\sim3-4$ \citep[see e.g.][]{Holden14}. We will discuss models that will provide
greater insights into these strong emission line sources in a future paper.

Our confirmation of a source with extremely strong rest-frame optical emission lines at $z_\mathrm{spec}=7.730$ together with two very similar sources at $z_\mathrm{spec}=7.213$ and $7.508$ \citep{Ono12,Finkelstein13} provides further support for the likelihood of ubiquitous strong rest-frame optical lines as evidenced in the IRAC photometry of $z\sim7-8$ galaxies. This has important consequences for future observations with JWST, which promises extremely efficient spectroscopic follow-up of such strong line emitters with NIRspec out to the highest redshifts of currently known galaxies.

%\vspace*{0.5cm}

\acknowledgments{
The authors thank the referee, Eros Vanzella, for very helpful feedback to improve this paper.
This work was supported by NASA grant NAG5-7697 and NASA grants HST-GO-11563.01, HST-GO-13792. 
RS acknowledges the support of the Leverhulme Trust.
The authors wish to recognize and
  acknowledge the very significant cultural role and reverence that
  the summit of Mauna Kea has always had within the indigenous
  Hawaiian community. We are most fortunate to have the opportunity
  to conduct observations from this mountain.
  This work is in part based on data obtained with the \textit{Hubble Space Telescope} 
  operated by AURA, Inc. for NASA under contract NAS5-26555, and with the Spitzer Space Telescope, operated by the Jet Propulsion Laboratory, California Institute of Technology under NASA contract 1407.
}

{\it Facilities:} \facility{Keck:I (MOSFIRE)}, \facility{HST (ACS, WFC3)}, \facility{Spitzer (IRAC)}

%\bibliography{MasterBiblio}
%\bibliographystyle{apj}

\end{document}